# Understanding the swelling behavior of P(DMAA-co-MABP) copolymer in paper-based actuators

Catarina C. Ribeiro[1,a,*], Nele Link[2,b,*], Jan-Lukas Schäfer[2,c], Carina Breuer[2,d], Markus Biesalski[2,e], Robert W. Stark[1,f,†]

[1] Institute of Materials Science, Technical University of Darmstadt, Peter-Grünberg-Str. 16, 64206 Darmstadt, Germany
[2] Department of Chemistry, Macromolecular Chemistry & Paper Chemistry, Technical University of Darmstadt, Peter-Grünberg-Str. 8, 64287 Darmstadt, Germany

[a] catarina.ribeiro@tu-darmstadt.de

[b] nele.link@tu-darmstadt.de

[c] jan-lukas.schaefer@posteo.de

[d] carina.breuer@tu-darmstadt.de

[e] markus.biesalski@tu-darmstadt.de

[f] robert.stark@tu-darmstadt.de

*both authors contributed equally to this work*

[†]*corresponding author*

# Abstract

As interest in sustainable materials grows, paper is being reimagined as a multifunctional substrate with significant potential for future technologies for innovative, environmentally friendly solutions. This study investigates the swelling behavior and environmental responsiveness of a copolymer, poly(N,N-dimethylacrylamide-co-4-methacryloyloxybenzophenone) [P(DMAA-co-MABP)], when applied to cellulosic paper for use in humidity-sensitive actuators. The copolymer's swelling behavior was characterized using dynamic vapor sorption (DVS) and in-situ atomic force microscopy (AFM). DVS measurements demonstrated that the polymer coating significantly enhances the hygroscopic properties of the paper, while AFM revealed the polymer's fast response to relative humidity (RH) changes, shown by immediate height adjustments, increased adhesion, and decreased stiffness at higher RH levels. Studies on polymer-modified paper-based bilayer actuators demonstrate that incorporating the hydrophilic P(DMAA-co-MABP) results in actuation in response to relative humidity variations between 10% and 90% RH. From these findings, two models were proposed to assess key mechanisms in the swelling behavior: the correlation between the heterogeneity in crosslinking and the polymer swelling behavior, and the correlation between polymer-paper interactions and the hygro-responsive bending behavior. Additionally, thermal analysis was performed by thermogravimetric analysis (TGA) and differential scanning calorimetry (DSC), providing a comprehensive profile of the copolymer's behavior.

*Hygro-responsive, Biomimetics, Photo-crosslinking, Polymer Swelling, Nanomechanical properties, P(DMAA-co-MABP), Bilayer, Actuators*

# Introduction

Paper has emerged as a versatile platform in advanced technological fields due to its accessibility, biodegradability, and biocompatibility. Innovations have enabled paper to serve as a responsive substrate in fields such as flexible electronics (Sala de Medeiros et al. 2020; Xu et al. 2021; Kumar et al. 2024), microfluidics (Yetisen et al. 2013; Akyazi et al. 2018; Hillscher et al. 2021), actuators (Wang et al. 2018; Poppinga et al. 2021), and sensor technology (Liana et al. 2012; Tai et al. 2020), fields that were traditionally dominated by synthetic, rigid substrates. Inspired by nature, biomimetics seeks to design materials that can passively adapt and respond to external stimuli (Ren et al. 2021). Many natural systems exhibit passive movements, where structures bend or deform in response to environmental changes. By functionalizing paper with responsive materials, researchers are developing devices that replicate these mechanisms by responding to environmental stimuli, such as humidity, light, temperature, electricity, magnetism, or fluids (Tai et al. 2020; Liu et al. 2021). For example, coating paper with humidity-responsive materials allows it to exhibit controlled shape changes, such as bending or curling, in response to moisture (Poppinga et al. 2021). These developments hold significant promise for eco-friendly devices in areas like soft robotics, medicine, and architecture (Le et al. 2019; Ren et al. 2021).

In a recent study inspired by motile plants like *Selaginella lepidophylia* (Rafsanjani et al. 2015), Schäfer et al. (2023) created polymer-modified paper sheets capable of hygro-responsive bending, mimicking the natural actuation seen in these plants. This study introduces a novel design for single-layer paper-based actuators with a polymer gradient across the paper thickness, resulting in increased dry and wet strength while enabling hygro-responsive movement. By manipulating drying conditions, they achieved directed polymer transport in the fiber network, allowing the paper to function as a self-contained bilayer, where the modified and unmodified layers interact in a complementary manner to induce bending. This phenomenon is analogous to the bimetallic beam bending behavior under thermal load, a well-known phenomenon from classical mechanics (Young and Budynas 1989). A bimetallic beam consists of two layers with different thermal expansion coefficients. It experiences differential strain when the temperature increases, which induces a bending moment due to the unequal expansion of the two layers.

Schäfer et al. used the copolymer P(DMAA-co-MABP) to achieve the desired hygro-responsive bending properties in the polymer-modified paper sheets. The polymer consists of two key components: *N,N*-dimethylacrylamide (DMAA), which imparts hydrophilic characteristics, and 4-methacryloyloxy-benzophenone (MABP), which contains photoreactive benzophenone groups. The MABP is particularly significant as it enables the copolymer to undergo photochemical reactions, allowing for stable cross-linking upon UV exposure. When activated by UV light, the benzophenone groups in MABP transition to a reactive triplet state, initiating C,H insertion cross-linking with adjacent polymer chains and fibers in the paper matrix (Toomey et al. 2004). This creates stable covalent bonds that firmly secure the polymer within the paper network. In addition to its application as an actuator agent, the P(DMAA-co-MABP) copolymer has also been used in other studies as a wet-strength agent (Jocher et al. 2015; Janko et al. 2015; Schäfer et al. 2021), protein-repellent coating (Meinzer et al. 2022; Bentley et al. 2022), patterned coating for microfluidics (Böhm et al. 2013), and as a humidity sensor (Kelb et al. 2016).

Nevertheless, to our knowledge, the work from Toomey et al. (2004) represents one of the few studies investigating the swelling behavior of P(DMAA-co-MABP) polymer networks. Their study focused on comparing thin, surface-attached P(DMAA-co-MABP) copolymer networks to bulk, non-attached configurations. Surface-attached polymer networks, covalently bonded to a substrate, exhibited limited swelling freedom, with expansion restricted largely to one dimension perpendicular to the surface. As a result, these networks show significantly lower in-plane hygroexpansion than bulk polymers of similar crosslinking density. Using multiple-angle ellipsometry, the study reveals that surface confinement significantly influences swelling characteristics of P(DMAA-co-MABP) networks, with results that qualitatively support an adapted Flory-Rehner theory for one-dimensional swelling. These findings highlight the distinct swelling and structural behavior that arise in surface-attached networks.

Understanding the polymer's swelling characteristics and its impact on the hygroexpansion of paper is essential for optimizing its applications in paper-based actuators and other moisture-sensitive materials. This study aims to comprehensively examine the swelling behavior of P(DMAA-co-MABP) copolymer in response to varying humidity conditions. Our investigation primarily focused on the polymer's hygroscopic properties, which were assessed using dynamic vapor sorption (DVS) to characterize its moisture uptake behavior. Furthermore, we conducted in-situ atomic force microscopy (AFM) measurements on surface-attached polymer thin films to gain insights into the polymer's hygroscopic and mechanical responses to humidity. These measurements allowed us to track real-time changes in the polymer's swelling and mechanical properties as a function of relative humidity (RH). To demonstrate the effect of the polymer on the hygroexpansion behavior of polymer-coated papers, we conducted actuation experiments using bilayers composed of polymer-modified paper strips. Based on our observations, we proposed two mechanisms for the polymer's response to moisture, addressing how heterogeneity in crosslinking affects the swelling behavior of the polymer and how the polymer plays a role in enhancing the hygro-responsive bending behavior of paper-based actuators. Additionally, we performed thermal analyses, including thermogravimetric analysis (TGA) and differential scanning calorimetry (DSC), to evaluate the polymer's thermal stability and transition behavior. Together, these analyses provide an in-depth understanding of the polymer's structural and functional behavior under dynamic environmental conditions.

## Materials and Methods

### a) Materials

**Materials.** The chemicals and solvents used, including DMAA, MABP, dimethylformamide (DMF), AIBN, diethyl ether, chloroform, and isopropanol (IPA), were purchased from Merck (Rove, NJ, USA), TCI (Eschborn, Germany), Alfa Aesar (Haverhill, MA, USA), Alberdingk Boley (Greensboro, NC, USA), Fisher Scientific (Hampton, NH, USA), Fluka Honeywell (Selze, Germany), Covestro (Leverkusen, Germany), and TIB Chemicals (Mannheim, Germany).

**Copolymer synthesis.** The polymer P(DMAA-co-MABP) was synthesized according to the procedure of Toomey et al. (2004) via radical copolymerization of DMAA and MABP. The commercially obtained DMAA was purified and destabilized by mild distillation at 90 °C under vacuum. For copolymer synthesis, DMAA (388.18 mmol, 98.07 mol%), MABP (7.63 mmol, 1.93 mol%), and AIBN (1.52 mmol, 0.38 mol%) were used. Under a nitrogen atmosphere, inhibitor-free DMAA was added to a 500 mL Schlenk flask containing MABP dissolved in 200 mL DMF. After adding AIBN as the initiator, the solution was degassed three times using the freeze-pump-thaw method. The reaction was carried out under a nitrogen atmosphere in an oil bath at 60 °C for 20 h. To stop the reaction, the solution was precipitated in multiple steps by adding approximately 15 mL of the copolymer solution in DMF at a time into 800 mL of diethyl ether. This stepwise precipitation was intended to reduce the possibility of agglomeration and to yield a finer copolymer powder for easier further processing. The resulting precipitate was collected by filtration. The off-white precipitate was dried under vacuum for over 48 h at 40 °C. Subsequently, the product was redissolved in 150 mL chloroform (CHCl$_3$) and again stepwise reprecipitated from cold diethyl ether (ratio 1:30), resulting in a white powder, which was further dried at 40 °C for 72 h. $^1$H-NMR was used to examine the chemical structure and determine the monomers ratio, resulting in about 98 mol% of the matrix DMAA and 2 mol% of MABP (see supplementary Fig. S1 and Calc. 1). The final copolymer, with a number-average molecular weight of 31,000 g/mol (Đ ~ 4.0), as determined by size exclusion chromatography (SEC), showed no detectable impurities or residual monomers according to NMR analysis (see supplementary Fig. S2). The copolymer powder was stored in a plastic container in the fridge until further use.

**Paper sheets.** Laboratory paper sheets were prepared from cotton linters pulp following the Rapid-Köthen process in accordance with DIN EN ISO 5269-2, with a grammage of 79 g/m$^2$ and free of any additives or fillers. Paper samples for DVS measurements were cut into pieces of 1×1 cm and into 1.5×12 cm stripes for actuator studies.

## b) Methods

**Dynamic vapor sorption (DVS):**

**Sample preparation.** Three types of samples, polymer foil, uncoated paper, and polymer-coated paper, were prepared to study the polymer's moisture sorption behavior and influence on cellulosic papers.

For the polymer film sample, the copolymer P(DMAA-co-MABP) was dissolved in a solution of 80% deionized water and 20% isopropyl alcohol (IPA) at a concentration of 175 mg/mL and stirred for 48 h at 350 rpm in a tightly sealed glass bottle. This concentration was selected to produce films with a thickness that enables complete detachment from the substrate. 2 mL of the copolymer solution was applied onto a Teflon plate using an Eppendorf pipette, and the polymer was crosslinked with a UV light wavelength of 365 nm and an energy dose of 16 J/cm$^2$ for 20 min, creating a foil that was later detached from the Teflon substrate.

No preparation was needed for the three uncoated paper samples.

For the polymer-coated samples, the copolymer P(DMAA-co-MABP) was dissolved in deionized water at a concentration of 50 mg/mL. According to the work of Schäfer et al. (2023), this concentration results in approximately 15 wt% polymer content within the paper. The paper samples were coated via a dip-coating process. The samples were immersed in the copolymer solution for 10 min, then dried on a nylon net with minimum contact points with the sample to reduce the risk of non-uniform solvent evaporation. After drying for 24 h under exclusion of light, the samples were exposed to UV light with an energy dose of 16 J/cm$^2$ for 20 min in order to crosslink the polymer with itself and with the paper network.

**Analysis.** All samples, eight in total (two polymer films, three uncoated papers, and three polymer-coated samples), were tested with a DVS analyzer (proUmid GmbH & Co. KG, Ulm, Germany) at 25 °C. The RH was increased stepwise by steps of 10%, from 0% RH until reaching 90% RH and back to 0% RH. Each step was held until the sample mass reached equilibrium.

**Atomic force microscopy (AFM):**

**Sample preparation.** A thin polymer film was prepared on a glass substrate using a dip-coating process, where the glass was immersed in a 50 mg/mL P(DMAA-co-MABP) copolymer solution in deionized water. After drying for 24 h under exclusion of light, the sample was exposed to UV light with a wavelength of 365 nm and an energy dose of 16 J/cm$^2$ for 30 min to crosslink the polymer. Finally, a sharp edge was created by carefully scratching the film with a needle to allow for thin film analysis.

**Analysis.** All measurements were performed using a Dimension Icon atomic force microscope (Bruker AXS, Santa Barbara, CA, USA) and a HQ:NSC35/AI-c cantilever (µMash, Sofia, Bulgaria) with a force constant of $k = 13.8$ N/m as determined with the thermal noise method (Butt and Jaschke 1995). A climate chamber was integrated into the AFM setup, allowing the adjustment of the RH according to the need to study the influence of RH in-situ.
In the PeakForce tapping mode, the slow scan axis was disabled so that the measurement was performed at the same line consecutively, a method we termed "one-line mode". Measurements were conducted on the sharp edge of the polymer. Here, three images were taken consecutively on a 20 µm line with 1008 samples/line, a tip velocity of 20 µm/s at a constant peak force of 25 nN and an excitation frequency of 2 kHz (100 nm oscillation amplitude). While the images were being taken, the RH was cycled between 70% and 90% for 5.5 cycles. These images were stitched and analyzed with the Gwyddion software, with cross-sectional analysis performed across height, adhesion, dissipation, and Young's modulus channels, averaging each section with 128 adjacent cross-sections. Simultaneous topographical and mechanical analyses were also performed over a 20 µm$^2$ area (with 256 samples/line) by enabling the slow scan axis, using a tip velocity of 20 µm/s, a peak force of 25 nN, and an excitation frequency of 2 kHz. The topography images were corrected with a first-order flattening filter using the NanoScope Analysis 1.9 software from Bruker.

**Actuator studies of P(DMAA-co-MABP)-modified paper bilayer:**

**Sample preparation.** The paper samples were coated via a dip-coating process. For this purpose, the copolymer P(DMAA-co-MABP) was dissolved in IPA at a concentration of 44 mg/mL and the paper strips were immersed in the solution for 1 min. For optimal comparability with the DVS samples, the paper strips were then dried on a nylon net with minimal contact points to reduce the risk of non-uniform solvent evaporation. After drying for 1 h in the absence of light, the samples were exposed to UV light at a wavelength of 365 nm with an energy dose of 16 J/cm$^2$ for 20 min to crosslink the polymer both, with itself and with the paper network. To remove non-crosslinked polymer residues, the samples were washed in three cycles of 30 min each in cold deionized water. The samples were then dried overnight under standard climate conditions. The polymer content in the paper samples was determined to be 16 wt%.

To form the paper actuators with an "active" and a "passive" side, bilayers were constructed by laminating a modified paper strip to a non-modified strip of the same size. Both strips were wetted by dipping them into a reservoir filled with deionized water, then placed on top of each other and gently pressed together. The side of the modified strip that faced upward during all modification steps also faced upward during lamination. The two strips were then placed between two sheets of non-stick paper (baking paper) and transferred to the drying section of a Rapid-Köthen sheet former. They were dried for 10 min at 90 °C at 900 mbar. Until further use, the samples were stored under standard climate conditions.

Illustrations of the sample preparation can be found in supplementary Fig. S6.

**Experimental Setup & Analysis.** The experimental setup for the actuator studies consisted of a chamber constructed from acrylic glass, which was connected to a humidity generator (MHG32TC, ProUmid GmbH & Co. KG, Ulm, Germany). The humidity inside the chamber was monitored by a sensor. The actuation of the paper samples was measured over two humidity cycles, with a total duration of 41 h. During these cycles, the relative humidity was alternated between 90 % and 10 %. Each measurement cycle began at a relative humidity of 50 %. The humidity was first increased to 90 % in two incremental steps, followed by a reduction to 10 % in four steps. Each step was held for 2 h, while the plateau phases at 90 % and 10 % relative humidity were held for 4 hours to ensure that the target humidity was achieved. The cycle concluded by increasing the humidity from 10 % back to 50 % (see Fig. 1).

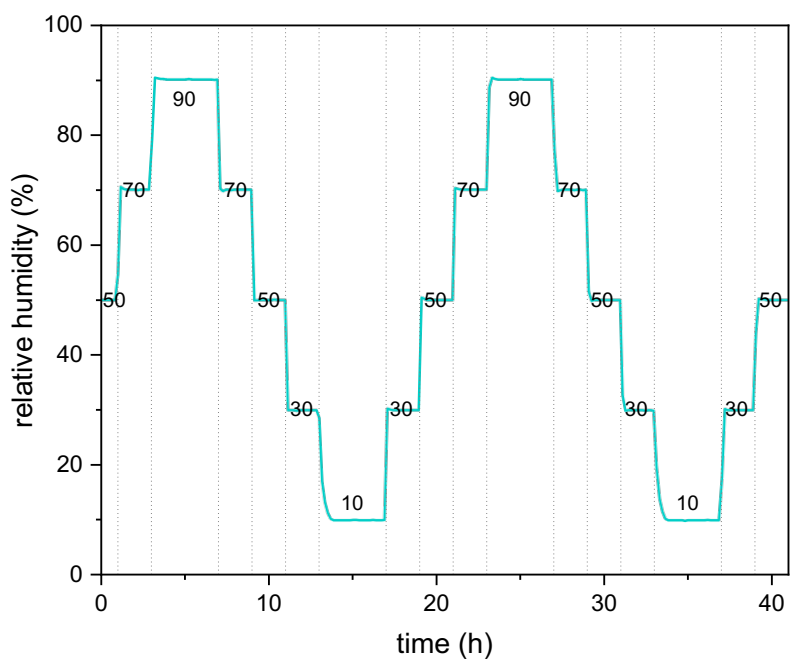

**Fig. 1** Humidity protocol for actuator studies: Two cycles each starting at 50% RH, increasing the relative humidity to 90% RH in two steps, decreasing the humidity to 10% RH in four steps and increasing again to 50% RH in 2 steps. Dashed vertical lines indicate the change of humidity setpoint.

During the measurements, the bending was monitored from above using a camera (Canon EOS 600D) mounted on a tripod, which captured an image every 10 min. Each bilayer was held in place by a clamp, with the polymer-modified side oriented to the left (see Fig. 2).

The movement of the bilayers was analysed automatically using Python software (Python 3.12). Each image was converted to grayscale and filtered to reduce background noise. The position of each bilayer was determined by identifying the white pixels with the largest and smallest *y*-values. The bilayer deflection was then calculated by subtracting the *x*-values of these two points and referencing the result to the initial deflection at the start of the experiment.

By correlating the timestamps of the images with the humidity data, the deflection of the strip was linked to the corresponding relative humidity levels.

In total, two sets of measurements were conducted, each using two bilayers.

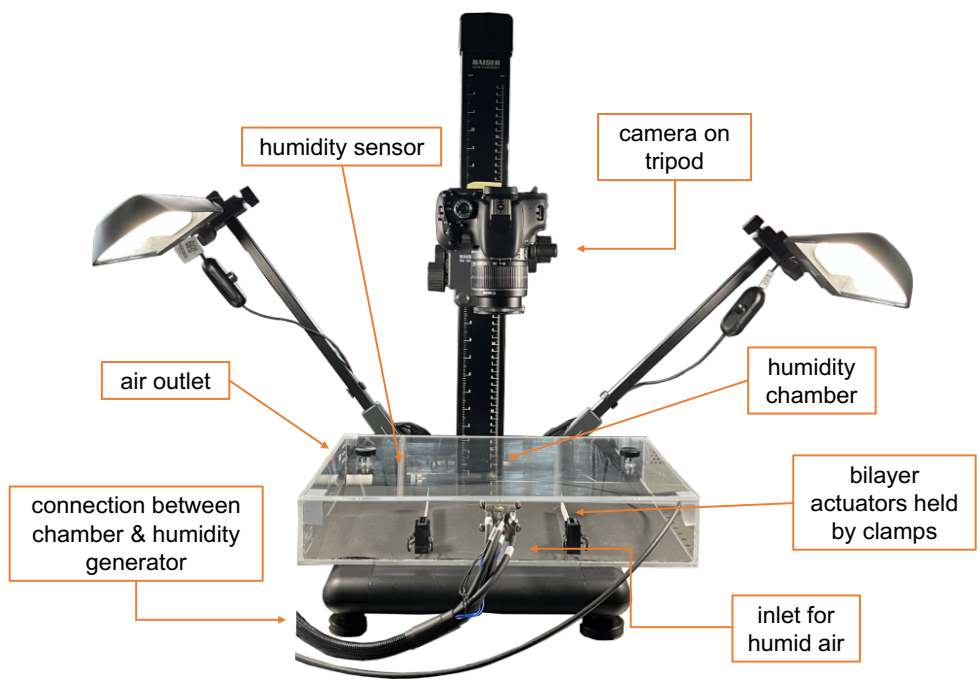

**Fig. 2** Experimental Setup for actuator studies consisting of humidity chamber with two bilayer actuators and a camera on a tripod.

**Thermogravimetric analysis.** A thermogravimetric analysis of the P(DMAA-co-MABP) copolymer foil, the coated, and the uncoated paper was carried out using a Thermogravimetric analyzer type TGA 1 (Mettler Toledo, Columbus, OH, USA) under a nitrogen atmosphere, with a heating rate of 10 °C/min over a temperature range of 25 °C to 600 °C.

**Differential scanning calorimetry.** The glass transition temperature of P(DMAA-co-MABP) copolymer foil was determined using a differential scanning calorimeter type DSC 3 (Mettler Toledo, Columbus, OH, USA) equipped with a FRS 5+ sensor. Two heating-cooling cycles were performed with a heating rate of 10 °C/min under a nitrogen atmosphere, over a temperature range from -50 °C to 300 °C. The second heating cycle was used for the determination of the glass temperature. The midpoint method, as defined by ISO 11357-2, was applied to determine the glass temperature. Differential scanning calorimetry was also conducted on the coated and the uncoated paper with the same heating rate (10 °C/min) and nitrogen atmosphere, over a temperature range from -75 °C to 250 °C for two heating-cooling cycles.

## Results and discussion

### a) Hygroscopic behavior of the P(DMAA-co-MABP) copolymer and cellulosic paper with DVS

In this work, paper was functionalized with the copolymer P(DMAA-co-MABP). The chemical structure of the copolymer is shown in Fig. 3a). The copolymer was attached to the paper fibers via the benzophenone groups by UV light treatment. This crosslinking reaction is shown schematically in Fig. 3b). The chemical reaction mechanism is shown in supplementary Fig. S1.

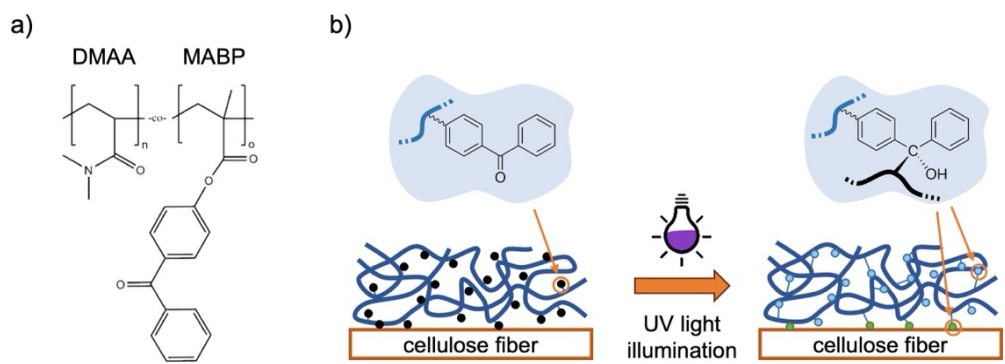

**Fig. 3** a) Molecular structure of P(DMAA-co-MABP) copolymer. b) Illustration of the photo-crosslinking reaction: benzophenone-containing units are represented as black dots. Upon exposure to UV light, these units form covalent bonds to cellulose fibers (green dots) and within the polymer matrix (blue dots).

To assess the swelling behavior of the P(DMAA-co-MABP) copolymer in humid air and its impact on the moisture absorption of the cellulosic paper when coated, DVS isotherms were obtained for uncoated paper, polymer-coated paper, and polymer film samples. The resulting sorption isotherms (Fig. 4) show distinct moisture uptake for each sample.

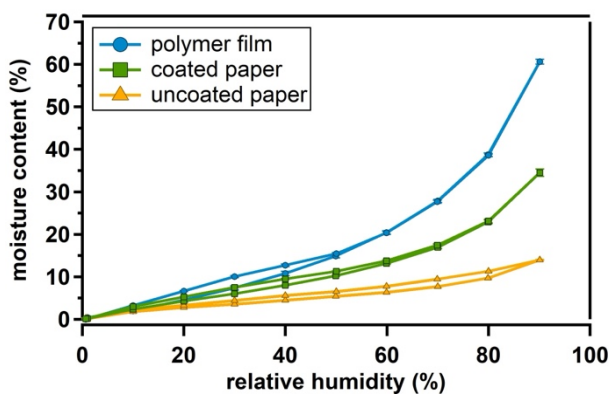

**Fig. 4** Sorption and desorption isotherms of uncoated paper, polymer-coated paper, and polymer film samples. Error bars represent standard deviation ($n = 3$), most are smaller than the symbol size.

All moisture sorption isotherm curves exhibit an S-shaped curve, which follow a type II isotherm. According to Brunauer's classification (Brunauer 1934), this shape is typical for materials with high moisture capacity at high RH, where the initial monolayer adsorption is followed by multilayer adsorption and capillary condensation.

The uncoated paper exhibited a gradual increase in moisture content with increasing RH, reaching a maximum of 13.9 ± 0.2% at 90% RH. This behavior is typical for cellulosic materials, which gradually absorb water as hydrogen sites become more available in the cellulose structure with increasing humidity (Ramarao and Chatterjee 2018). A low hysteresis over the full RH range was observed, often attributed to fiber swelling during adsorption and irreversible pore closure due to hydrogen bond formation during desorption, which can restrict complete moisture release from the fibers (Kulachenko 2021).

The polymer film absorbed significantly more water, 60.6 ± 0.6% moisture at 90% RH, with a notably sharper increase after 50% RH, reflecting the polymer's strong water absorption capacity. The second part of the isotherm (RH>50%) not only shows a steep increase in moisture content but also exhibits little to no hysteresis, both of which suggest that at higher RH, water uptake is dominated by multilayer adsorption, capillary filling and bulk swelling, which are processes more reversible in nature. In contrast, the hysteresis observed in the low RH range (<50%) likely results from stronger water-polymer interactions, such as hydrogen bonding, structural arrangements, or water entrapment within the polymer network, which restrict desorption.

The polymer-coated paper showed a distinct isotherm compared to the uncoated paper, with enhanced moisture uptake particularly above 50% RH, reaching 34.5 ± 0.9% at 90% RH, 2.5 times greater than that of the uncoated paper. The sharper uptake and reduced hysteresis at higher RH suggest that the polymer coating dominates the hygroscopic response of the polymer-coated paper, increasing its moisture absorption capacity driven by the polymer's high affinity for water.

### b) In-situ AFM analysis of polymer thin film at different RH conditions

In-situ one-line mode AFM measurements were performed to investigate the polymer behavior with the change of RH. Therefore, the RH was cycled between 70% and 90% for 5.5 cycles, as seen in Fig. 5a). Fig. 5b) shows the AFM height image that resulted from varying the RH. Note that, as the slow scan axis was disabled, the image does not show the topography of the thin film but rather the changes in height of a single line over time. The dark region is the surface of the substrate whose height stays constant over time while varying the RH. The brighter region is the surface of the polymer thin film, and it can notably be seen that its height changes with varying RH by observing the changes in brightness over time. To further investigate these changes, the cross-section of the height over time at the purple line shown in Fig. 5b) was taken and averaged over the 128 adjacent points represented by the faded purple area. The same was done for the channels of adhesion, dissipated energy, and Young's modulus. These cross-sections are shown in Fig. 5c). Note that the cross-section of the height was normalized relative to its initial height (486 nm) at 5.7% RH. The time scales of Fig. 5a-c) are the same, therefore, we can observe the immediate changes in the polymer thin film with varying RH. Fig. 5d) presents both the 2D and 3D representation of the setup, while Fig. 5e) displays two representative force-distance curves at 70% and 90% RH.

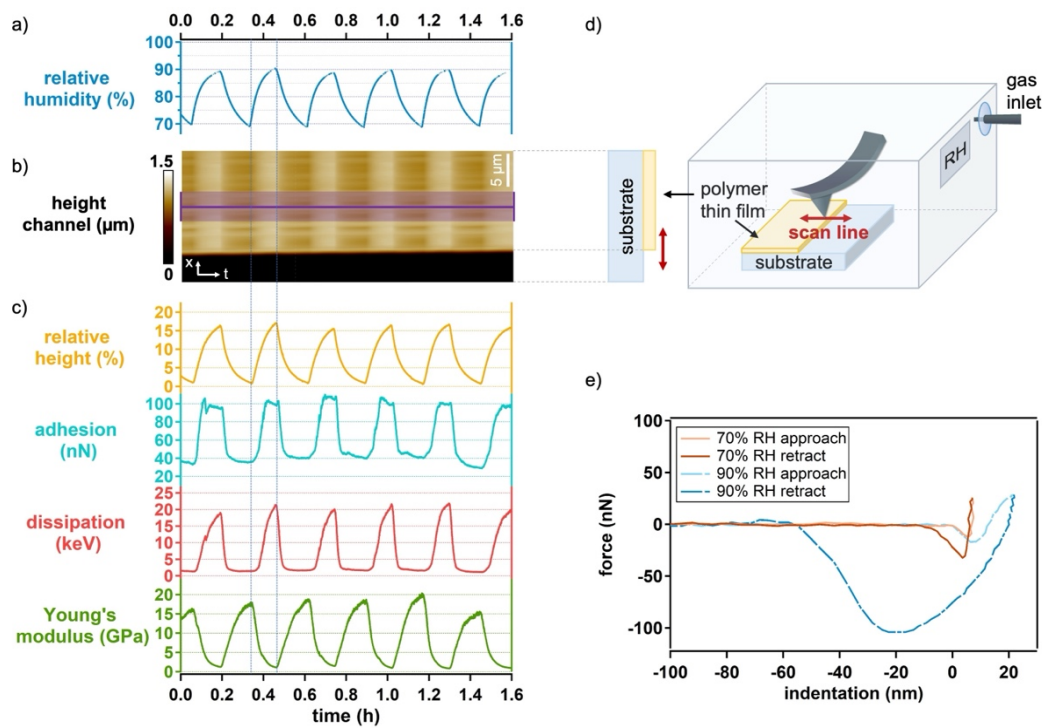

**Fig. 5** In-situ AFM study on the humidity-responsive behavior of the P(DMAA-co-MABP) copolymer thin film. a) Changes in relative humidity plotted over time. b) AFM height image from one-line mode measurement (fast scan axis, x-axis, against time, t-axis). The purple line represents the cross-section taken for the analysis and averaged over the 128 adjacent points, represented by the faded purple area. c) polymer properties extracted from the purple cross-section: relative height, adhesion, dissipation, and Young's modulus. d) 2D and 3D illustration of the one-line experiment set-up. e) Two representative force-distance curves at 70% and 90% RH.

The experimental results reveal a strong correlation between the RH and the physical properties of the polymer. Notably, the polymer's height responds almost immediately to changes in RH, increasing and decreasing proportionally as RH rises and falls, respectively. This indicates a rapid absorption of moisture by the polymer, which results in its swelling. Adhesion forces between the polymer and the cantilever tip also significantly increase in response to higher RH levels, with adhesion values spiking up to 100 nN. This sharp increase in adhesion indicates a stronger interaction between the polymer surface and the cantilever tip, likely due to a change in the viscosity of the polymer and the formation of a water meniscus or enhanced capillary forces at higher RH levels. This change in adhesion is accompanied by an increase in energy dissipation, which occurs with a time delay relative to the initial rise in RH, suggesting that the polymer requires a certain amount of time or a threshold level of water uptake to change. Additionally, the stiffness derived from the force-distance curves shows a decrease as RH increases, with Young's modulus values dropping from a range of 15-20 GPa at 70% RH down to 2 GPa at 90% RH. This is a 90% reduction of the polymer's initial stiffness, which reflects the polymer's softening as it swells with absorbed water, altering its mechanical properties. It is reasonable to correlate the polymer stiffness with its molecular structure and the interactions between polymer chains and water molecules. As water diffuses into the polymer matrix, water molecules weaken the intermolecular forces (e.g., hydrogen bonds and van der Waals forces) between polymer chains, leading to increased chain mobility. This, consequently, causes the polymer network to expand, resulting in swelling. This expansion, combined with the loosening of the polymer structure, compromises the polymer's mechanical integrity, leading to a significant reduction in stiffness. In contrast, in a dry state, polymer chains are closely packed, which contributes to a higher stiffness.

In sum, the AFM results demonstrate that the polymer is highly responsive to changes in RH, as evidenced by the immediate height changes, increased adhesion, and decreased stiffness as RH increases. These results highlight the polymer's sensitivity to environmental humidity, making it suitable for applications where responsiveness to moisture is crucial.

Upon reviewing the bending behavior in paper-based actuators, we noticed a key difference between our results and those from Schäfer et al. (2023). Over seven cycles of RH change, Schäfer et al. observed that while the actuator's bending at low humidities remained nearly constant with minimal attenuation, its ability to return to the initial position at high humidity progressively decreased with each cycle. While no definitive conclusions were drawn, they outlined several potential factors for this attenuation, based on similar reported behaviors: reduced moisture absorption due to morphological changes in the paper (e.g., hornification) (Larsson and Wågberg 2008), plastic deformation caused by bending at high humidity levels (Kulachenko 2021), and changes in the polymer's swelling behavior after multiple cycles.

In contrast, our AFM results, which were conducted on the polymer alone, show no significant variations in height or elastic modulus over multiple RH cycles, indicating that the polymer does not exhibit ageing effects. Since our results were focused exclusively on the polymer, it is reasonable to attribute the decrease in deflection observed by Schäfer et al. to ageing within the paper's fiber network rather than the polymer.

To better understand the role of the paper substrate, we conducted AFM and DVS measurements on uncoated paper samples subjected to six RH cycles. The AFM results revealed no significant changes in Young's modulus across the cycles (see Supplementary Fig. S4), indicating stable mechanical properties. In contrast, DVS measurements showed a gradual decrease in moisture uptake with each cycle, with an overall reduction of 5.2% after the sixth cycle (Supplementary Fig. S5). These findings suggest that, although the paper shows no significant degradation of its mechanical integrity after the studied cycles, the paper's moisture absorption capacity slightly decreases, possibly due to morphological changes as previously mentioned.

Nevertheless, while the morphological changes in paper reduce its moisture absorption and consequently its hygroexpansion, this alone does not explain the actuator's diminished ability to return to its initial position observed by Schäfer et al. In fact, a decrease in the paper's hygroexpansion would theoretically favor bending back toward the uncoated paper layer, promoting the recovery to its initial position rather than restricting it. Therefore, the observed loss in reversibility suggests that additional mechanisms, such as plastic deformation of the paper and/or accumulative residual stresses, may play a more dominant role.

To further study the influence of RH on the polymer's topographical and mechanical properties, AFM measurements were performed on the polymer film at two distinct states: dry (25% RH) and swollen (95% RH), as shown in Fig. 6.

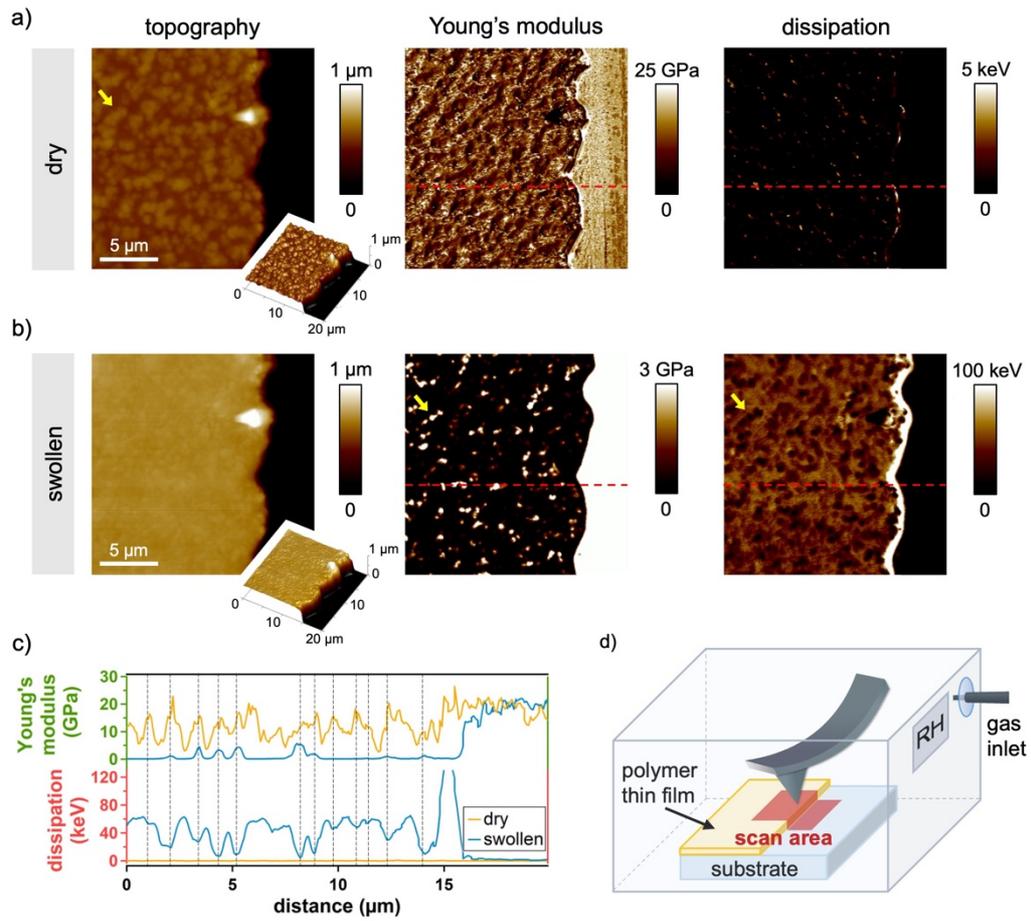

**Fig. 6** 2D and 3D topography, Young's modulus, and dissipation AFM images of the P(DMAA-co-MABP) polymer thin film at a) dry (25% RH) and b) swollen (95% RH) states. c) Young's modulus and dissipation cross-section from the red dashed line depicted in a) and b). d) 3D illustration of the experiment set-up.

At 25% RH, the polymer surface exhibited a rather rough surface, an average Young's modulus of 12 GPa and dissipated energy of 0.3 keV, indicating a relatively stiff material with minimal viscoelastic behavior. In contrast, at 95% RH, the polymer revealed a swollen and smoother surface, its Young's modulus dropped to an average of 0.4 GPa, and the dissipated energy increased significantly to 40 keV, which matches the previous results from the one-line mode AFM experiments. Notably, several regions across the polymer surface maintained a higher stiffness (3-6 GPa) and lower dissipated energy (6 keV) in the swollen state. These regions align with the round areas observed in the topography image in the dry state, which are pointed out in Fig. 6a and b) with yellow arrows. The cross-section of the Young's modulus and dissipation taken from the red-dashed line, plotted in Fig. 6c), reveals that the regions that retained a higher stiffness and lower dissipation at the swollen state exhibited an initial higher Young's modulus at the dry state. Furthermore, these regions have an average diameter of approximately 0.75 µm, however, considering dilation effects from tip convolution, their actual size is likely smaller (García 2010). In a previous study by Bentley et al., AFM topographic images of P(DMAA-co-5%MABP) spray coated on a PMMA substrate showed similar surface features with an average 0.5 µm size (Bentley et al. 2022). In their study, they did not conclude the origin of the features, however, it is reasonable to suggest that the observed features are likely of the same nature.

### c) Correlation between polymer swelling behavior and crosslinking densities

We propose the model shown in Fig. 7 to explain the appearance of the morphological features of the polymer films and their response to swelling. Upon UV irradiation, benzophenone groups in MABP enter a reactive triplet state and abstract hydrogen atoms from nearby C–H bonds on adjacent polymer chains. The resulting radicals recombine to form covalent C–C bonds, establishing stable crosslinks within the polymer network (Toomey et al. 2004). Given the low MABP content (2 mol%) and the mechanism of UV-induced crosslinking, together with the AFM topography and mechanical maps (Fig. 6), we suggest that the crosslinking density varies locally within the polymer matrix, as illustrated in Fig. 7 (light and dark areas representing low and high local crosslinking density regions, respectively). This may arise from the random distribution of MABP units along the polymer chains, combined with variations in local polymer concentration during solvent evaporation in the film formation and drying process.

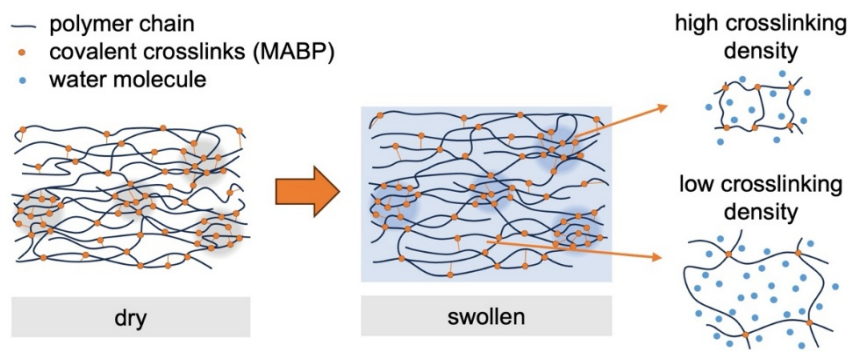

**Fig. 7** Illustration of the polymer matrix in the dry and swollen state and the interaction between high and low local crosslinking density regions with water molecules: The darker areas correspond to regions with higher crosslinking density. Upon water uptake, the high crosslinking density regions

exhibit limited swelling and largely retain their original conformation, whereas the low crosslinking density regions allow greater chain mobility and expand significantly, resulting in more pronounced configurational changes and loss of mechanical stability.

This heterogeneity in local crosslinking density is consistent with our AFM observations, which show regions with different mechanical stiffness. In the dry state, both high and low crosslinking density regions maintain compact conformation. Upon swelling, regions with low crosslinking density have more space between polymer chains, allowing greater freedom for water molecules to infiltrate and expand the polymer network, resulting in increased chain mobility and a softer structure. In contrast, high crosslinking density regions have a tightly interconnected network that resists water penetration and limits swelling. These regions would remain comparatively stiffer in a swollen state as the dense network structure resists expansion and retains its mechanical integrity.

### d) Actuator studies of polymer-modified paper bilayers

According to Schäfer et al. (2023), the bending behavior observed in polymer-modified paper actuators is primarily driven by the differences in the hygro-responsiveness of the paper and the polymer. They attributed the bending of the actuator to stress from drying, differential swelling and compaction between the unmodified "resistance" layer and the polymer-modified "active" layer (Fig. 8). Bending is caused by a misfit strain caused by differential hygroexpansion within the paper-based actuator under varying humidity conditions. The fact that there is a different concentration of polymer within the bilayer leads to internal stresses and strains and, finally, to changes in the shape of the actuator.

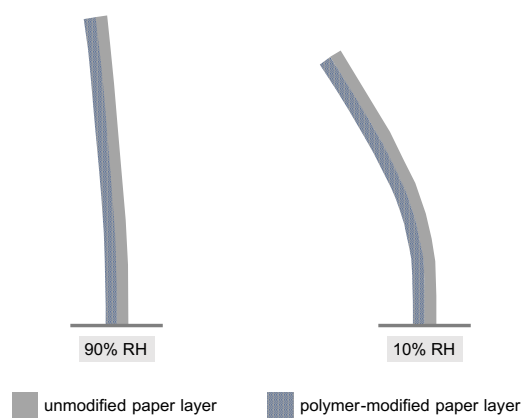

**Fig. 8** Illustration of the bending effect of a paper-based bilayer actuator (consisting of an unmodified "resistance" layer and a polymer-modified "active" layer) with deflection at 90 % and 10 % RH.

Fig. 9 illustrates the movement of a representative polymer-modified, paper-based bilayer actuator during two cycles of varying humidity. The images show the actuator at 50%, 90%, and 10% relative humidity. After an initial increase in humidity from 50% to 90% RH, the actuator bends toward the polymer-modified, "active" side of the bilayer as the humidity decreases, and bends back into its initial position as the humidity increases again.

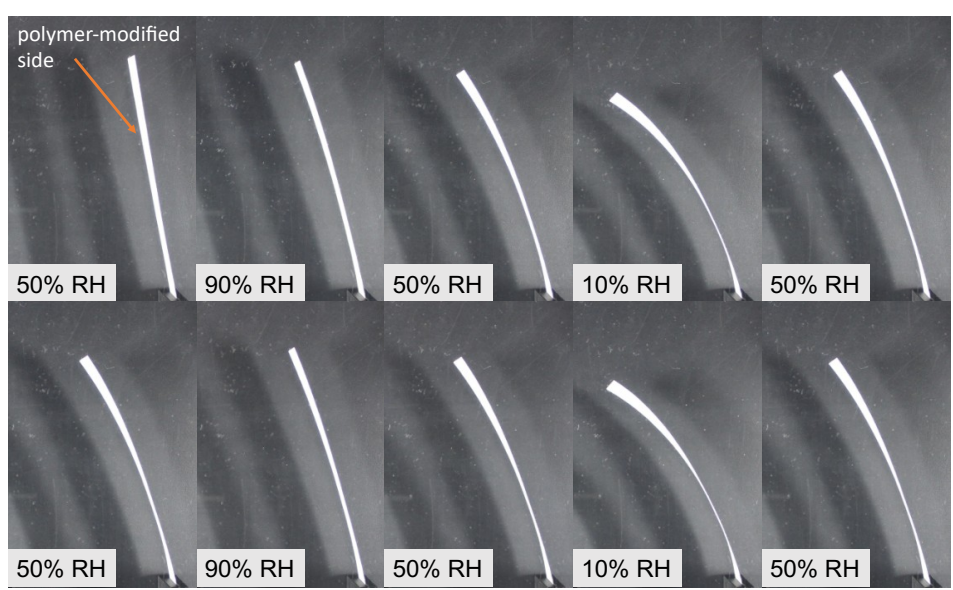

**Fig. 9** Sequential photographs of a polymer-modified paper-based bilayer actuator subjected to two humidity cycles ranging from 90% to 10% RH (unmodified "resistance" layer facing right and a polymer-modified "active" layer facing left). To improve visualization of the bending mechanism, the final image of the top row (50% RH) and the initial image of the bottom row are intentionally repeated.

Fig. 10 shows the average deflection of four actuators during two humidity cycles. During the first cycle, the actuators bend toward the polymer-modified side as the humidity increases from 50% to 90% RH, reaching a mean deflection of -5.3 mm. In four steps, the relative humidity is then decreased to 10% RH, resulting in a deflection of -32.6 mm. This corresponds to a total bending of 27.3 mm. In the second humidity cycle, the actuators bend back toward their initial state, reaching a deflection of -6.7 mm at 90% RH. Upon a subsequent decrease in humidity, they again bend 26 mm, ending at a deflection of -32.7 mm. These results show that while the deflection at low relative humidity remains nearly unchanged, the actuators do not return fully to their original position at higher humidity levels. These findings are consistent with those reported by Schäfer et al.

To verify that the actuator movement is indeed influenced by the polymer-modified paper, we also conducted control experiments using bilayers composed of two non-modified paper strips. These bilayers showed no significant deflection in response to changes in humidity (see Supplementary Fig. S7).

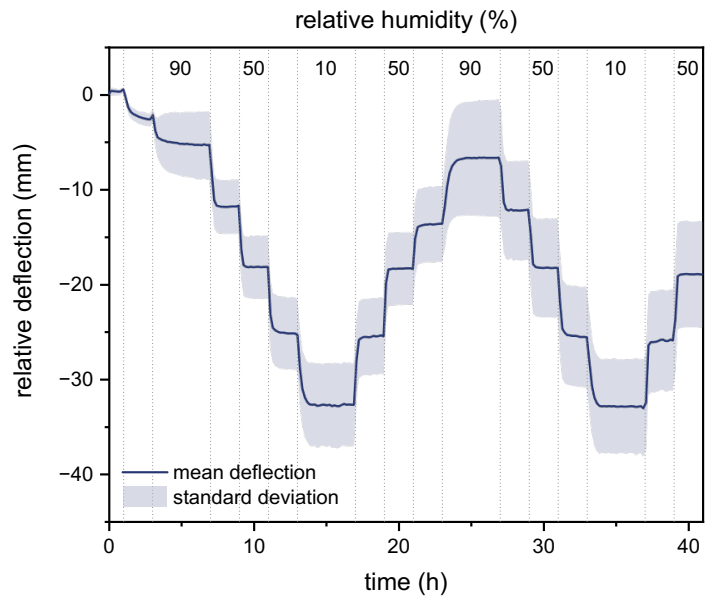

**Fig. 10** Average deflection of four actuators during two humidity cycles: mean deflection shown in blue, the standard deviation from four samples shown in grey and areas of different relative humidity illustrated by dashed lines.

e) Correlation between polymer-paper interaction and hygro-responsive bending behavior

In the following, we discuss a model, built on the previous works of Schäfer et al. and Toomey et al. (2004), that considers the hygro-responsiveness of both the paper and the polymer, as well as polymer-paper unique interaction.

Brandberg et al. (2020) demonstrated that in-plane hygroexpansion, despite its smaller magnitude compared to out-of-plane hygroexpansion, is the primary driver of macroscopic bending in paper. However, while out-of-plane hygroexpansion of paper ranges from 20% to 30% (Tydeman et al. 1965), in-plane hygroexpansion is considerably smaller, with reported values around 0.5% to 0.9% (Larsson and Wågberg 2008; Viguié et al. 2011; Manninen et al. 2011).

As for the polymer, insights from Toomey et al. (2004) provide further context to the hygroexpansion of P(DMAA-co-MABP) copolymer: in semi-confined, surface-attached polymers, swelling predominantly occurs perpendicular to the surface (out-of-plane direction) due to restrictions on lateral movement (in-plane directions), resulting in the buildup of in-plane stresses (Fig. 11a). AFM measurements show that the polymer exhibits out-of-plane hygroexpansion of up to 20%. Although the polymer in-plane hygroexpansion was not determined, it is expected to be significantly lower, based on the work of Toomey et al..

To extend Toomey et al.'s model to the paper-based actuator herein, it is necessary to take into account the structural complexity of the fibrous paper network, particularly at the inter-fiber bonding regions (Fig. 11b). Unlike a flat substrate, paper consists of an entangled, interwoven fiber network, where fiber orientation is highly heterogeneous (at least for lab sheets used here) (Fig. 11c and d). Because the polymer is covalently attached to the fibers, the polymer hygroexpansion is not isotropic, but it is guided by fiber orientation and bonding regions. Therefore, we suggest that at the inter-fiber bonding regions, where fibers are most strongly interconnected, the geometry allows the polymer to undergo multidirectional expansion (Fig. 11b). This enhances the in-plane hygroexpansion of the polymer-coated paper layers. Brandberg et al. (2020) demonstrated that in-plane hygroexpansion originates at inter-fiber bonds, supporting our hypothesis that these regions play a crucial role in in-plane hygroexpansion and that the presence of the polymer in these regions is expected to further amplify this effect. In addition, the polymer likely increases the effective Young's modulus of the composite material because the pore space of the fiber is filled with the polymer.

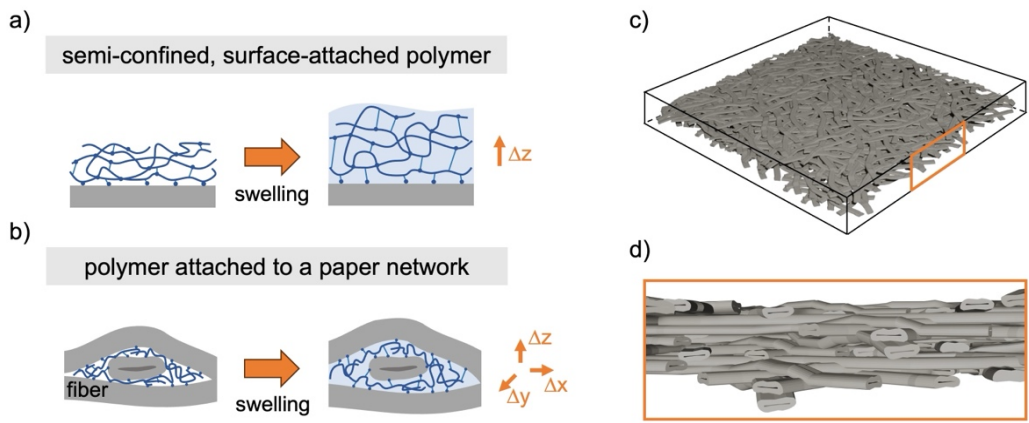

**Fig. 11** Illustration of the swelling behavior of a) a surface-attached polymer, where expansion is restricted to a single direction perpendicular to the surface due to confinement, b) a polymer coating within a paper sheet, where the polymer can swell in multiple directions due to the random

orientation of fibers in the paper network, c) 3D representation of a paper sheet and d) its profile view.

These mechanisms help to better understand the distinctive bending behavior of these polymer modified paper-based actuators, in which the differential in-plane hygroexpansion of the "active" and "passive" layers and differences in the mechanical properties contribute to the bending response. The mismatch in swelling between the two layers generates an in-plane stress σ. The difference in the hygroexpansion coefficients and the Young's moduli of the materials leads to different linear deformations between the coated and uncoated paper. This mismatch causes an in-plane deformation, leading to a curvature in the actuator. Mechanically, this is equivalent to an axial load $P = σw$, where $w$ is the width of the actuator. Thus, the strain of the paper actuator is determined by its effective Young's modulus $E^*$, its moment of inertia $I$, and the axial stress $P$. The effective Young's modulus $E^*$ has to account for the Young's moduli and thickness of both layers, as well as their position relative to the neutral axis of the bilayer system.

Stoney's equation is widely used to quantify the surface stresses generated in, for example, molecular thin films or bimetal actuators (Stoney 1909). Here, we can use this approach to make a first guess about the relevant parameters in the bilayer actuator. Nevertheless, one should keep in mind that one important assumption is that one layer is thin compared to the thickness of the actuator, which implies that the misfit strain is compensated within the coated layer. Within this approach, the vertical displacement of the actuator $v(x)$ can then be determined from the governing equation of beam deflection

$$E^*I \frac{d^4 v(x)_{axial}}{dx^4} - P \frac{d^2 v(x)_{axial}}{dx^2} = 0.$$

This equation accounts for both the bending stiffness and the axial load effects, where the axial load $P$ incorporates the in-plane stresses. For a regular cantilever, the moment of inertia is $I = \frac{wt^3}{12}$, where $w$ is the width, and $t$ is the thickness of the cantilever.

The governing equation of beam deflection highlights three relevant parameters that determine the shape of the bent actuator and that are affected by the polymer and by the moisture: the effective Young's modulus $E^*$, the moment of inertia $I$ and finally, the surface stress which leads to an axial load $P$. Thus, we can identify the following mechanisms that are relevant to better understanding the bending: changes in the Young's modulus due to moisture, changes in the cross-section due to swelling and the equivalent axial load due to differential swelling. The first and the second effect affect the mechanical stiffness of the actuator, whereas the differential swelling leads to in-plane stress and thus bending.

### f) Thermal characterization of the P(DMAA-co-MABP) copolymer with DSC and TGA

To verify the thermal stability of the system under study, we performed thermogravimetric analysis (TGA) and differential scanning calorimetry (DSC) on the uncoated paper substrate, the polymer-coated paper composite, and the P(DMAA-co-MABP) copolymer (Fig. 12).

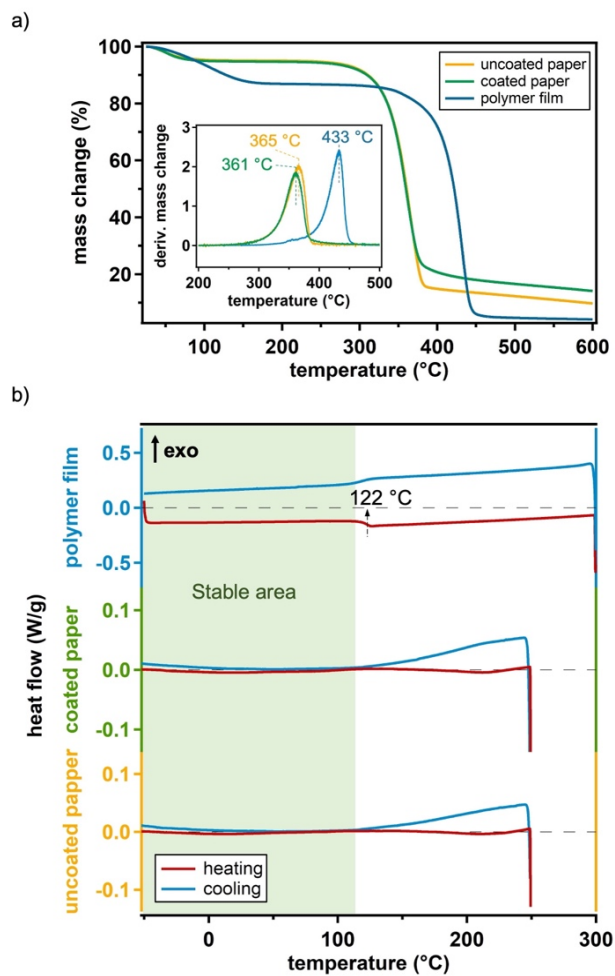

**Fig. 12** a) Thermogravimetric analysis and its derivative, and b) differential scanning calorimetry curves of uncoated paper, polymer-coated paper, and P(DMAA-co-MABP) polymer film.

TGA results revealed a two-step thermal degradation process for all samples (Fig. 12a): the first degradation phase, occurring between 30-175 °C, exhibited a mass loss of 4.9%, 5.3%, and 13.2% for the uncoated paper, the coated paper, and the copolymer film, respectively. This initial mass loss is attributed to the evaporation of residual water and is consistent with the DVS data. The second degradation step showed substantial mass losses of 85.3%, 80.6%, and 82.6%, with degradation rates peaking at 365 °C, 361 °C, and 433 °C for the uncoated paper, the coated paper and the copolymer film, respectively. This second step signals thermal decomposition of the polymer and paper matrix. Both the coated and uncoated papers exhibited similar thermal degradation, with decomposition occurring at lower temperatures than the polymer film. However, the coated paper showed a higher residual mass compared to the uncoated sample, which is expected due to the presence of the polymer, as it does not fully decompose. A previous study by Hamou and Djadoun reported a similar two-step degradation in PDMAA polymer and claimed that the second step is ruled by the main chain degradation and by decarboxylation and carbonization processes (Hadj Hamou and Djadoun 2011).

The polymer film DSC curves (Fig. 12b) revealed an endothermic transition characterized by a baseline offset, without any crystallization and melting peaks. This behavior suggests an amorphous polymer structure with a glass transition of 122 °C, supporting the mechanism we hypothesize for the polymer matrix structure. In contrast, no significant differences were observed between the DSC curves of the uncoated and polymer-coated papers. Notably, the glass transition of the polymer was not detectable in the coated samples, which could be explained by the polymer being covalently bound to the paper fibers, which restricts its molecular mobility.

Overall, the thermal analyses confirm that all materials remain stable well below 100 °C (illustrated as a green area in Fig. 12b), which is the relevant temperature range for the paper-based actuator operation. Since no thermal transitions occur in this range, thermal effects can be considered negligible during operation.

## Conclusion

This study advances the understanding of the hygro-responsive behavior of P(DMAA-co-MABP) copolymer when used in paper-based actuators, with a focus on the polymer swelling behavior. DVS and AFM measurements demonstrated the rapid and reversible response to humidity changes, revealing significant moisture uptake, height increase, and stiffness reduction under high RH conditions. Studies on polymer-modified paper bilayer actuators demonstrate that incorporating the hydrophilic P(DMAA-co-MABP) results in actuation in response to relative humidity variations between 10% and 90% RH.

Two possible mechanisms for the response to moisture were discussed: the first is that the heterogeneity in crosslinking within the polymer network affects its swelling properties, with areas of stronger crosslinking potentially resisting expansion and remaining stiffer in humid conditions. While variations in crosslinking density were not experimentally studied, the impact on the polymer's overall structural integrity and responsiveness to moisture could be explored in future research. The second mechanism addresses the role of the polymer on the hygro-responsive bending behavior of polymer-modified paper actuators. Due to the covalent bonding of the polymer to the entangled interwoven fiber network, with particular emphasis on inter-fiber bonds, swelling occurs multi-directionally rather than only perpendicular to the paper surface, as expected from surface-attached polymers. This multi-directional expansion and contraction enhances in-plane strain. As bending in bilayer actuators arises from differential in-plane strain between the active and passive layers, the polymer-induced amplification of in-plane deformation is key to the bending behavior of the actuator in response to humidity.

Additionally, the study highlights key mechanisms influencing the bending behavior of the actuator, such as changes in Young's modulus due to moisture, swelling-induced changes in the cross-section, and the equivalent axial load resulting from differential swelling.

Thermal analysis, including TGA and DSC, further confirmed the stability of the copolymer and paper-based materials within the actuator's range. All components remained stable under 100 °C, with the copolymer exhibiting a glass transition temperature of 122 °C. Together, these findings provide a deeper understanding of the copolymer's functionality in paper-based actuators driven by changes in humidity.

**Acknowledgments**

The authors thank Liliya Dubyey for conducting DVS measurements.

## Statements and Declarations

**Funding.** This work was supported by the German Research Foundation (DFG) for financial funding under grant No. 40554961, grant No. 405422473 and grant No. 405422473, respectively.

**Competing Interests.** The authors declare no conflict of interest.

**Author Contribution.** C.C.R. and N.L. conceived and designed the study together with M.B. and R.W.S. C.C.R. conducted all AFM measurements. N. L. synthesized the P(DMAA-co-DMAA) polymer for the polymer analysis and bilayer formation and conducted the actuation experiments. J.L.S. synthesized the P(DMAA-co-DMAA) polymer for the DVS and AFM samples, while C. B. performed DCS and TGA measurements. The manuscript was written by C.C.R. and N.L. with support from J.L.S., M.B., and R.W.S. and further input from all authors. C.C.R. and N.L. prepared the figures. All authors read, reviewed, and approved the final manuscript. R.W.S. and M.B. supervised the project.